\pdfoutput=1

 \documentclass[final,3p,times,twocolumn]{elsarticle}

\usepackage{amssymb}

\usepackage{epsfig}
\usepackage{rotating}
\usepackage{amssymb}
\journal{Computer Physics Communications}

\begin{document}

\begin{frontmatter}



\title{A graphics processor-based intranuclear cascade and evaporation simulation}


\author{\corref{cor1}H.~Wan Chan Tseung}
\author{C.~Beltran}
\cortext[cor1]{Corresponding author. {\it Email}: wanchantseung.hok@mayo.edu}

\address{Division of Medical Physics, Department of Radiation Oncology, Mayo Clinic, Rochester MN 55905}

\begin{abstract}
Monte Carlo simulations of the transport of protons in human tissue have been deployed on graphics processing units  (GPUs) with impressive results. To provide a more complete treatment of non-elastic nuclear interactions in these simulations, we developed a fast intranuclear cascade-evaporation simulation for the GPU. This can be used to model non-elastic proton collisions on any therapeutically relevant nuclei at incident energies between 20 and 250 MeV. Predictions are in good agreement with Geant4.9.6p2. It takes approximately 2 s to calculate $1\times 10^6$ 200 MeV proton-$^{16}$O interactions on a NVIDIA GTX680 GPU. A  speed-up factor of $\sim$20 relative to one Intel i7-3820 core processor thread was achieved.
\end{abstract}


\end{frontmatter}


\section{Introduction}\label{}

Monte Carlo (MC) techniques are increasingly being used in proton therapy to simulate and validate treatment plans \cite{MGHMC, Beltran}. MC calculations are more accurate than pencil-beam evaluations, since they take into account detailed microscopic processes and have a better handling of material inhomogeneities in patients. However, computational times associated with MC simulations can be prohibitive. For example, calculating the dose within 2\% statistical error in a 150 cm$^3$ volume requires around 1 hour on a modern 100-node cluster\footnote{This simulation was carried out with TOPAS \cite{TOPAS}, a proton therapy simulation package that facilitates the use of Geant4 \cite{Geant4}.}. At present, the use of MC is therefore either restricted to institutions that have access to significant computing resources, or reserved for a handful of cases requiring detailed investigation and enhanced accuracy. 

Recently, there have been several attempts to simulate both charged particle and photon transport on graphics processing units (GPUs) \cite{Kohno, Hissoiny, Jahnke, Jia}. In particular, proton transport MCs for particle therapy applications have been successfully implemented. Using GPUs, MC calculations of treatment plans involving $1\times10^7$ proton trajectories have been completed in under 30 s. It is therefore clear that GPUs can be of tremendous benefit to proton therapy, both for research and in clinical applications (e.g. plan validation and optimization).

However, the current generation of GPU proton MCs make a number of simplifying assumptions concerning non-elastic nuclear interactions. Kohno et al.  \cite{Kohno} do not directly simulate nuclear processes, but `include' them by using measured proton depth dose distributions in water. Jia et al.  \cite{Jia} follow Fippel and Soukup \cite{Fippel} by using a crude model of nuclear interactions on the oxygen nucleus. They demonstrate that the lack of an in-depth model of nuclear processes does not significantly impact dose predictions in low-density tissue and bone. It is nonetheless conceivable that unsatisfactory results could be obtained in certain situations, e.g. propagation through high-Z materials such as metallic implants. Detailed investigations (e.g. studies of secondary particles and residual nuclei) are also not possible with the model used in refs. \cite{Jia, Fippel}.

To better include non-elastic
nuclear interactions, one can envisage a hybrid proton transport program, in which ionization energy loss, straggling and multiple scattering are simulated on the GPU, while nuclear events are handled by existing software packages on the CPU. However, in this configuration the running time is expected to be completely dominated by nuclear calculations. For $1\times10^7$ protons in the therapeutic energy range (70--250 MeV), we estimate the number of nuclear events that need to be simulated to be of the order of $10^6$. On a i7-3820 3.6 GHz processor it can take from 200 to 900 s to simulate $1\times10^6$ 200 MeV proton--$^{16}$O non-elastic interactions with Geant4.9.6p2, depending on the chosen nuclear interaction model.

Our primary aim is to improve the treatment of nuclear processes in a GPU proton transport MC without considerably extending the net calculation time. Thus, we developed a rudimentary but fast GPU-based MC simulation of nuclear interactions, which is able to predict secondary particle properties following non-elastic events for incident proton energies below 250 MeV. To our knowledge, such simulations have not been previously reported.  We show that on a NVIDIA GTX680 card, $1\times10^6$ 200 MeV  proton--$^{16}$O nuclear events can be computed in 2 s.

This paper is organized as follows. In \S2, we describe our approach, which is a conventional Bertini-type intranuclear cascade (INC) model including nuclear evaporation. We then give details of its implementation on the GPU in \S3. Our results are described in \S4. Predictions from our code are verified with Geant4.9.6p2; simulated data for proton interactions with $^{16}$O and $^{40}$Ca are shown, and calculation times are compared. We summarize in \S5.

\section{Non-elastic interaction model}\label{section:fullmodeldescription}
\subsection{Overview}
A non-elastic nucleon-nucleus interaction can be assumed to take place in two steps: a fast INC stage that leaves the nucleus in an excited state, followed by an evaporation stage. 
The INC phase has been modeled extensively since the late 1950's using MC techniques \cite{Metropolis, Bertini, Chen, G4Binary, G4Bertini}. In the Bertini approach \cite{Bertini}, the incident particle and nucleon-nucleon collision products are tracked, assuming straight line trajectories and taking into account medium effects, until they exit the nucleus or their energies lie below a cutoff. The simulated nucleon-nucleon collisions are not ordered in time. In the evaporation phase, de-excitation of the nucleus occurs after energy equilibration, first through the emission of low-energy particles (nucleons and possibly heavier fragments), then photons. The probability of particle emission can be calculated using statistical methods \cite{Weisskopf, Dostrovsky}.

The INC-evaporation model applies to non-elastic interactions in the therapeutic energy range down to a few tens of MeV, although its validity becomes questionable at low energy. It contains no adjustable parameters, i.e. its predictions do not need to be normalized. Our model inputs, assumptions and further details on the two phases are given below.

\subsection{Nuclear model}
The nucleus is treated as a two-component fermion gas mixture of protons and neutrons. The nucleon density $\rho(r)$ follows a Fermi distribution:
\begin{equation}\label{equation:fermidensity} \rho(r) \propto (1+\mathrm{e}^{\frac{r-c}{\delta}})^{-1}\end{equation}
where $c = 1.07A^{\frac{1}{3}}$ ($A$ is the mass number), and $\delta = 0.545$. For simplicity, $\rho(r)$ is approximated by a step function consisting of 15 shells of constant densities and equal thickness. The density of each shell is found by integrating $\rho(r)$.  Following Chen et al. \cite{Chen}, we take the radius of a nucleus to be $c+2.5$ fm. The potential energy $V$ of a nucleon is the sum of its Fermi energy $E_F$ and the binding energy $E_B$, which is calculated using the semi-empirical mass formula. Target nucleon momenta are sampled from a quadratic distribution.

\subsection{Intranuclear cascade}\label{section:cascadedescription}
As mentioned above, the incident nucleon and collision products are followed in the nucleus. The sequence of calculations in our cascade simulation roughly follows that of Metropolis et al. (fig.~3 in ref. \cite{Metropolis}).
Below 250 MeV, only nucleon--nucleon elastic collisions need to be considered.  Total and differential elastic nucleon-nucleon cross-sections are calculated using the parameterizations of  Cugnon et al. \cite{Cugnon}. Collision kinematics are fully relativistic.  Pauli blocking is enforced by requiring that the kinetic energies of all collision products exceed their Fermi energies. The cutoff energy is taken to be $E_F+E_B$ for neutrons and $E_F+E_B+E_C$ for protons, where $E_C$ is the Coulomb barrier. Refraction and reflection of nucleons at shell boundaries are taken into consideration using the prescription of Chen et al. \cite{Chen}. At the end of the cascade, the excitation energy of the residual nucleus is calculated through energy conservation,  as in ref. \cite{Metropolis}.

\subsection{Nucleon evaporation}
For the evaporation phase, we adopt the generalized evaporation model (GEM) \cite{GEM}. Although the GEM can handle nuclide emission up to Mg, only He and lighter particles are considered in this work. An isotropic angular distribution is assumed in the rest frame of the nucleus for all evaporated particles. Other post-cascade processes such as photon evaporation, pre-equilibrium emission of particles, fission and fermi break-up are not currently included in our model.

\section{GPU implementation}\label{section:implementation}
In this section, our implementation of the above model on a GPU\footnote{We assume the reader to be acquainted with GPU terminology; if not, one can refer to the abundant resources available on the NVIDIA developer website \cite{nvidia}.} using the CUDA framework \cite{cuda} is described.

\subsection{Software organization and memory allocation}
Three kernels are used for the following tasks: (1) initialization of random number sequences for generation within individual threads using the CURAND library \cite{curand}, (2) the INC simulation, and (3) nuclear evaporation calculations. Kernel 2 processes one full cascade and kernel 3 computes one nuclear de-excitation per GPU thread. 

As for memory allocation, GPU constant memory is used to store simulation-wide physics constants. Pre-calculated total nucleon-nucleon cross-section tables and other read-only input data arrays are stored as 1-D textures. Global memory is used to store information on incident particles, excited nuclei, and secondary particles that escape the nucleus. Collision products are stored in per-thread local memory before being simulated. Shared memory is minimally used in the present work. Floating point precision is adopted in all calculations. 

 Compile-time conditions are inserted to isolate CUDA C extensions, so that the same program can be made to run on the CPU. This greatly facilitates debugging and running time comparisons.


\subsection{Intranuclear cascade kernel}


The INC model described in \S\ref{section:cascadedescription} was implemented as one GPU kernel. The inputs to this kernel are: a structure of arrays (SoA) describing the incident protons (position, momentum, and energy) and the initial CURAND states for all threads. A SoA is used to ensure coalesced global memory reads. The outputs are two SoA's containing information on   exiting particles and the excited nuclei remaining after the INC. 

The limiting factors affecting run-time performance were: (1) thread divergence caused by the intrinsic stochastic nature of the simulation and conditional instructions, (2) register spill and  local memory overhead, due to the large kernel size and the use of local memory arrays to store secondary nucleons that need to be propagated in the INC.

To assess the impact of divergence, we simulated 70 and 200 MeV proton--$^{12}$C cascade events with all threads assigned the same initial random number seed (i.e. CURAND state). The running times were approximately four and five times shorter for the 70 and 200 MeV interactions, respectively. Thread divergence is unavoidable in MC simulations, but we tried to minimize its effects where possible, e.g. by avoiding deeply-nested conditional statements and factoring out operations that are common to all branches. 

Our attempts to store secondary nucleons in shared or global (instead of local) memory resulted in longer running times. Shared memory usage reduced GPU thread occupancy, and using global memory to keep track of recoiling nucleons proved to be comparatively less efficient due to very frequent uncoalesced access and stores (many secondary nucleons need to be processed in one cascade). 

\subsection{Particle evaporation kernel}
Our particle evaporation kernel is based on the Geant4.9.6p2 GEM code. To adapt the latter to the GPU, the object-oriented structure was discarded. Class methods used in calculating particle emission probabilities and sampling emitted particle kinetic energies were re-engineered as device functions.

The evaporation kernel inputs are the CURAND states and the SoA containing information on the post-INC excited nuclei. The output is another SoA describing the properties of evaporated particles. Since one thread can involve more than one exiting particle, the SoA is written to global memory in an uncoalesced manner. We verified that this did not result in a significant  increase in total running time. The limiting factors for the evaporation kernel were again thread divergence and register spill.


\section{Validation and results} \label{}

We calculated proton interactions on a number of therapeutically relevant nuclei at different incident proton energies between 70 and 230 MeV. To illustrate the performance of our MC, predicted secondary proton and neutron yields for proton--$^{16}$O and proton--$^{40}$Ca are shown in comparison with Geant4.9.6p2 simulations in \S\ref{section:g4comparison}. Computational times are then compared in \S\ref{section:runtime}.
 
\subsection{Comparisons with Geant4.9.6p2 and data}\label{section:g4comparison}
Geant4.9.6p2 calculations were performed using the binary and Bertini cascade models \cite{G4Binary, G4Bertini}, both of which are routinely adopted in Geant4-based proton therapy MCs. The binary cascade model is more detailed and computationally more intensive than the Bertini approach. Our code is conceptually closer to the latter. The two Geant4 models allowed us to gauge the level of discrepancy that can be expected among established INC simulations. 

\begin{figure}[]\centering
\includegraphics[width=8.6cm]{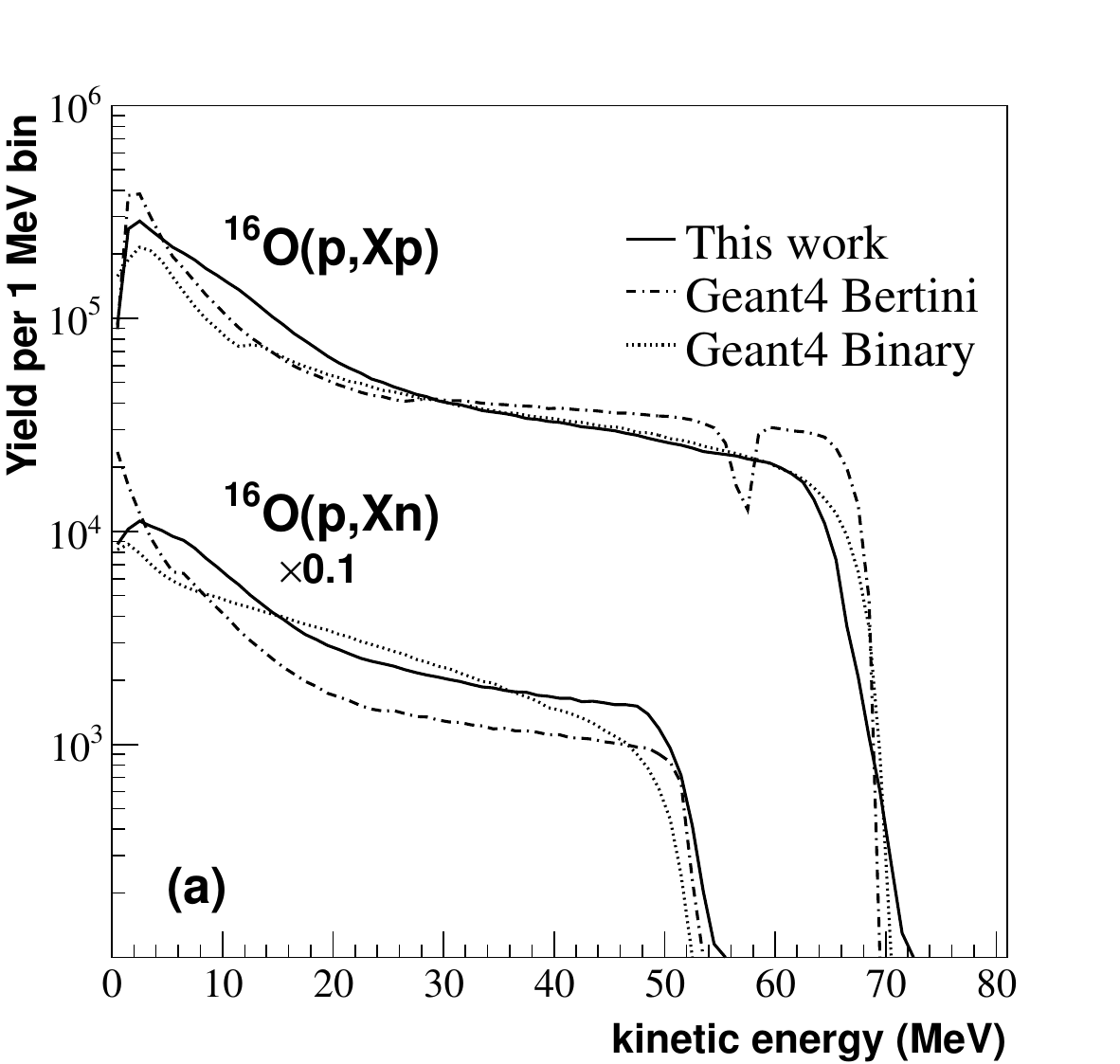}
\includegraphics[width=8.6cm]{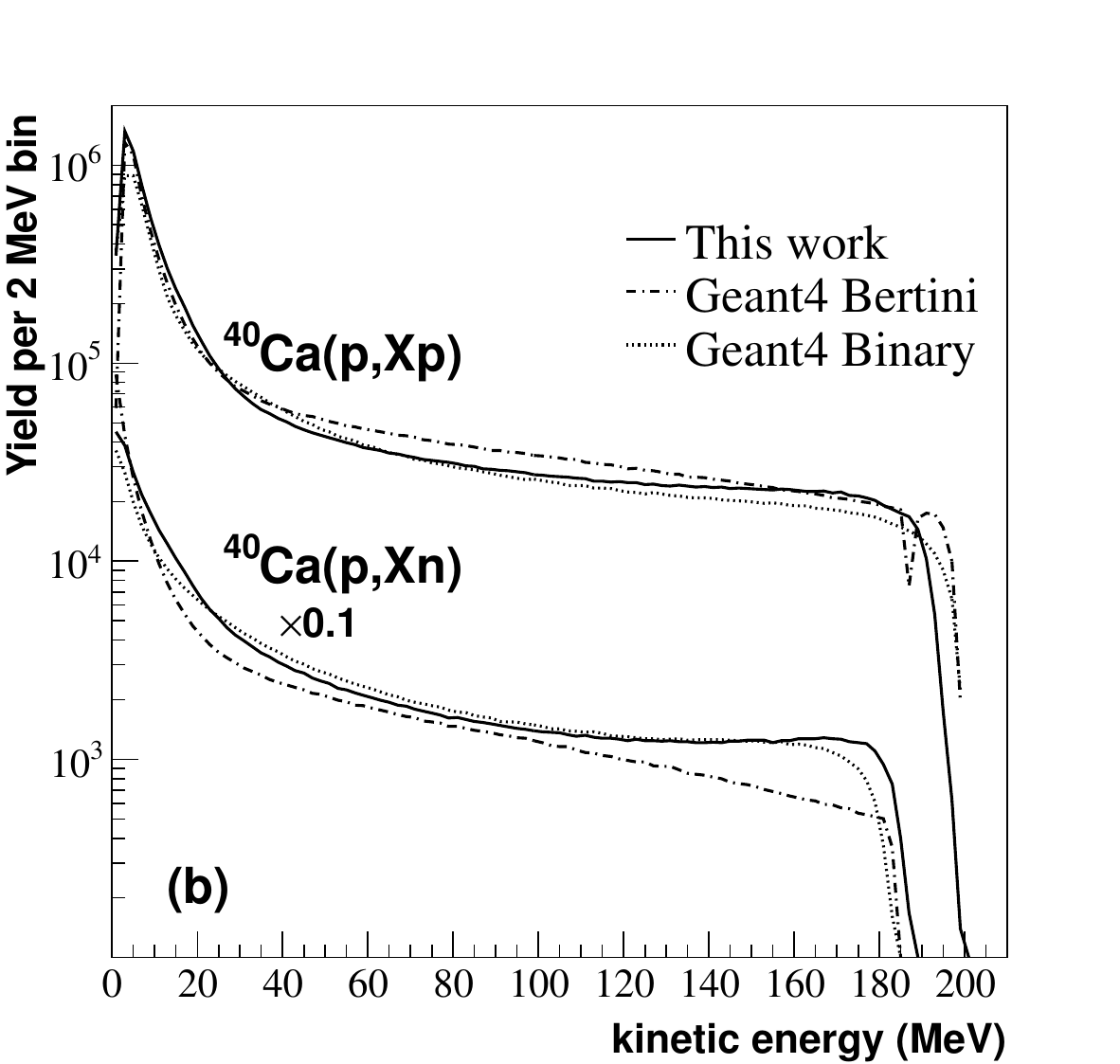}
\caption[]
 {Secondary proton and neutron yields as a function of energy for (a) 70 MeV proton--$^{16}$O and (b) 200 MeV proton--$^{40}$Ca interactions. }
 \label{figure:protonneutronyields}
 \end{figure}

\begin{figure}[]\centering
\includegraphics[width=8.6cm]{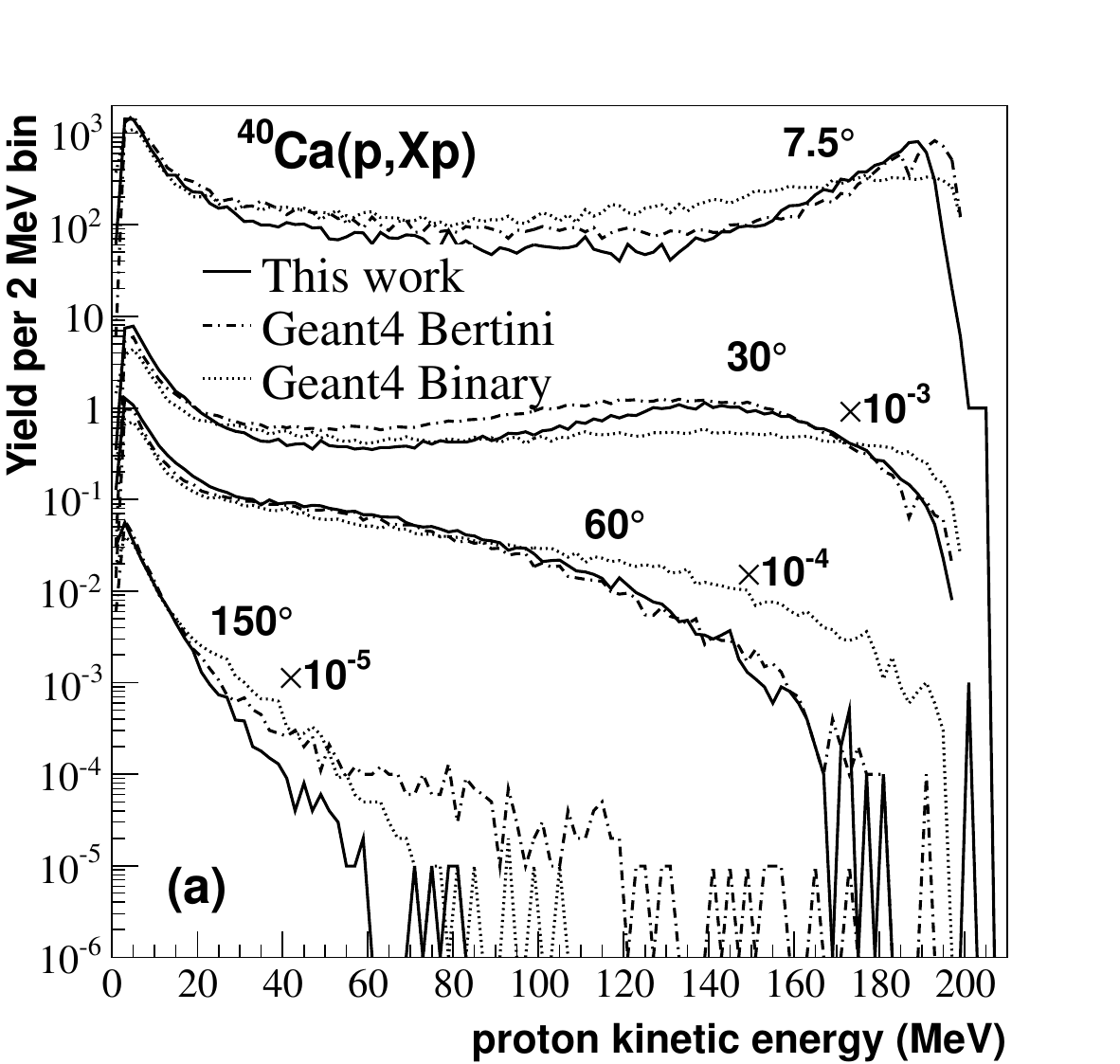}
\includegraphics[width=8.6cm]{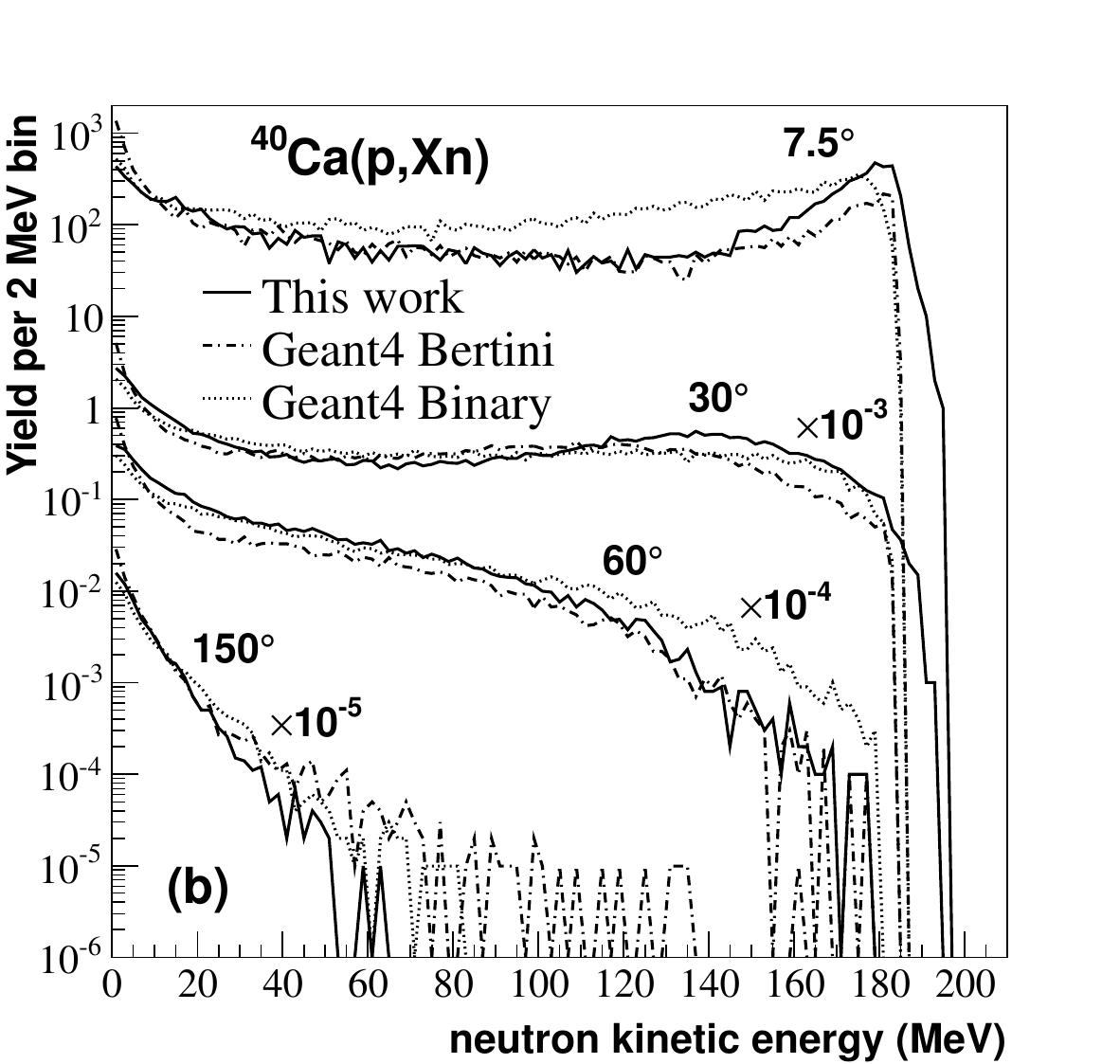}
\caption[]
 {Secondary (a) proton and (b) neutron yields in a $1^{\circ}$ bin centered at $7.5^{\circ}$,  $30^{\circ}$, $60^{\circ}$ and $150^{\circ}$, for 200 MeV non-elastic proton interactions on $^{40}$Ca.}
 \label{figure:protonneutrondiffyields}
 \end{figure}

\begin{figure}[]\centering
\includegraphics[width=8.6cm]{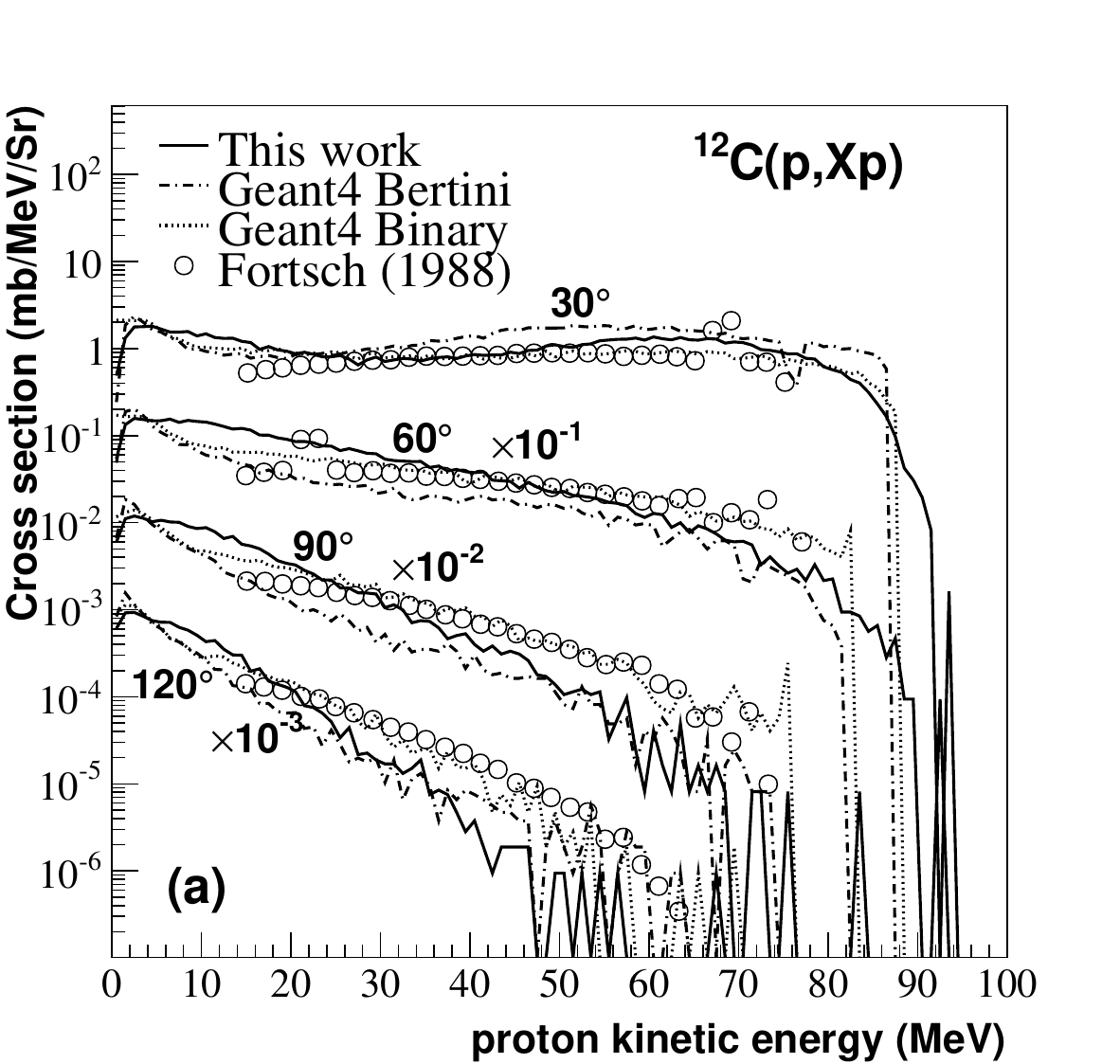}
\includegraphics[width=8.6cm]{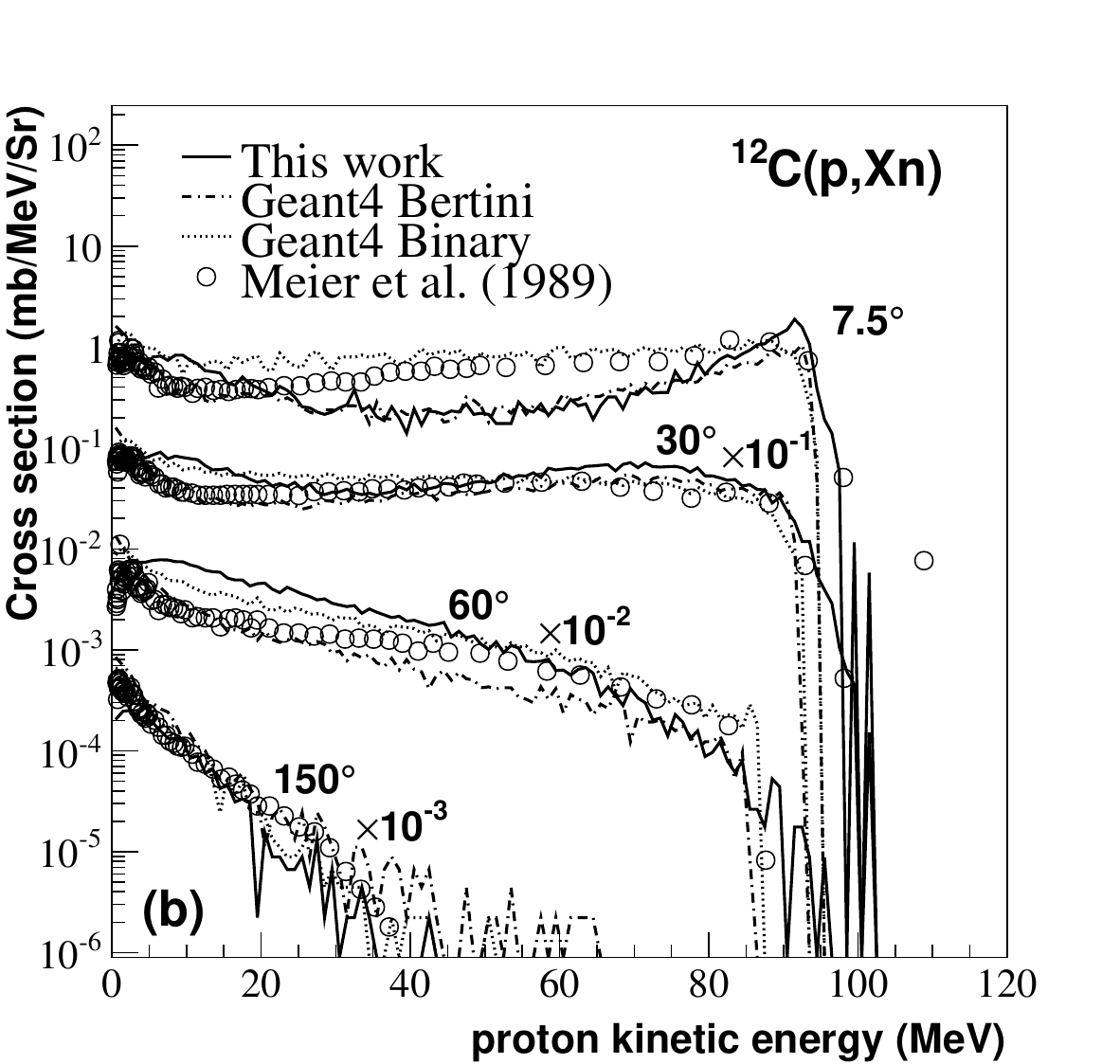}
\caption[]
 {Calculated secondary (a) proton and (b) neutron production cross-sections on $^{12}$C  compared with data, at 90 MeV and 113 MeV incident energies, respectively. Data from Fortsch et al. \cite{Fortsch} and Meier et al. \cite{Meier}.}
 \label{figure:diffyieldsvsdata}
 \end{figure}

Fig.~\ref{figure:protonneutronyields} shows our computed yield of secondary protons and neutrons for (a) 70 MeV proton--$^{16}$O and (b) 200 MeV proton--$^{40}$Ca interactions. The neutron yield has been scaled down by a factor of 10 for clarity.  Figs.~\ref{figure:protonneutrondiffyields}(a) and (b) show the proton and neutron yields in a $1^{\circ}$ bin centered at $7.5^{\circ}$,  $30^{\circ}$, $60^{\circ}$ and $150^{\circ}$ for 200 MeV proton--$^{40}$Ca interactions. The $30^{\circ}$, $60^{\circ}$ and $150^{\circ}$ curves have been  scaled down again for clarity. Fig.~\ref{figure:diffyieldsvsdata} shows the predicted secondary proton and neutron production cross-sections on $^{12}$C, at 90 and 113 MeV incident energies respectively. The measurements of Fortsch et al. \cite{Fortsch} and Meier et al. \cite{Meier}  are also shown for comparison.

The yields shown in figs.~\ref{figure:protonneutronyields} and \ref{figure:protonneutrondiffyields} are absolute, i.e. they were not normalized to each other. The GPU simulation is in reasonable agreement with both Geant4 models. Comparable levels of agreement with Geant4 were obtained for other therapeutically relevant nuclei (not shown here for brevity). 

\subsection{Run-time performance}\label{section:runtime}
\begin{table}[hpt]
\begin{center}
\footnotesize\rm
\begin{tabular}{@{}ccccc}
\hline Platform & Model & K.E (MeV)&p--$^{16}$O (s) &p--$^{40}$Ca (s)\\\hline\hline
(1) CPU& G4 Binary &70&2082.41&3220.15\\
(2) CPU& G4 Bertini &70&556.57&501.38\\
(3) CPU& see \S\ref{section:runtime} &70&138.21&242.94\\
(4) GPU& see \S\ref{section:runtime} &70&6.49&11.83\\
(5) GPU &see \S\ref{section:runtime}&70&5.04&7.19\\\hline
(1) CPU& G4 Binary &200&2382.07&4620.92\\
(2) CPU& G4 Bertini &200&492.42&534.13\\
(3) CPU &see \S\ref{section:runtime}&200&172.74&326.11\\
(4) GPU &see \S\ref{section:runtime}&200&8.39&15.56\\
(5) GPU &see \S\ref{section:runtime}&200&6.61&9.77\\\hline
\end{tabular}
\caption{\label{table:runtimeresults} Time taken in seconds to compute $2.6\times 10^6$ proton--$^{16}$O and proton--$^{40}$Ca interactions at 70 and 200 MeV incident proton kinetic energy (K.E). See \S\ref{section:runtime} for details.}
\end{center}
\end{table}
All simulations were preformed on a workstation equipped with a i7-3820 3.6 GHz processor and a GTX680 card with 4 Gb on-board memory. In table~\ref{table:runtimeresults}, we compare the calculation times for $2.6\times 10^6$ proton--$^{16}$O and proton--$^{40}$Ca  interactions at 70 and 200 MeV, for five cases: (1) Geant4.9.6p2 binary cascade model on the CPU, (2) Geant4.9.6p2 Bertini model on the CPU, (3) our MC compiled to run on the CPU, (4) our MC compiled to run on the GPU, and (5) our MC compiled with the \texttt {-use\_fast\_math} option to run on the GPU. This \texttt{nvcc} compiler flag enforces faster but less accurate CUDA intrinsic functions for some mathematical operations. CPU calculations were done on a single thread, and running times for (4) and (5) include host CPU operations. 

On the CPU, our MC is 2--4 times faster than the Geant4 Bertini cascade computations. On the GPU, our MC benefits from a speed-up factor of $\sim$20. Using CUDA math intrinsics, the performance is further improved speed-wise by a factor of 1.3--1.6. As far as we can tell, the resulting loss in accuracy in predicted secondary particle yields is small and its impact on dose calculations should be negligible.

For 200 MeV protons on $^{16}$O, the calculation time (as a percentage of the total run-time) of each kernel is as follows: 72\% for the INC kernel, 23\% for the evaporation kernel and 3\% for CURAND initialization. The rest of the time is spent in host-GPU data transfers. For 200 MeV protons on $^{40}$Ca the respective percentages are 50\%, 47\% and 2\%. The INC kernel run-time is roughly the same for $^{16}$O and $^{40}$Ca, while the evaporation kernel run-time is significantly higher for $^{40}$Ca. The run-times for heavier nuclei are expected to be dominated by evaporation calculations.

Although the run-time performance is adequate for our purposes, it can be improved in a number of ways. As seen in \S\ref{section:implementation}, our implementation is unsophisticated and not fully optimized for the GPU. To mitigate the effects of divergence, one can try to re-structure the code into small data-parallel tasks, i.e. separate kernels to handle vertices (e.g. shell boundaries  or  nucleon-nucleon interactions) and steps. Trivially, we might also benefit by using a compute capability 3.5 GPU, which allows four  times more registers per thread.


This INC-evaporation simulation is a critical part of our effort to develop a fast and accurate GPU-based particle transport MC for proton therapy dose calculations. From the results shown above, for $1\times 10^7$ proton trajectories the overhead due to non-elastic interactions is likely to be $< 10$ s on a GTX680. 
 Another potentially useful application of our code is a very fast calculation of the yield of position emitters for range verification. This is the subject of future work.

\section{Conclusions}
In summary, we have implemented an intranuclear cascade-evaporation simulation for the GPU. This simulation will result in more accurate and rigorous MC calculations of proton therapy treatment plans on the GPU. Our work also illustrates how GPUs can be used to significantly accelerate MC simulations of non-elastic nuclear interactions. 

\section{Acknowledgements}
This work was funded in part by a grant from Varian Medical Systems, Inc. We are grateful to our colleagues at Mayo Clinic for carefully reading the manuscript. 








\end{document}